\documentclass[twocolumn,showpacs,preprintnumbers,amsmath,amssymb
              ,prb,eqsecnum]{revtex4}

\usepackage{graphicx}
\usepackage{dcolumn}
\usepackage{bm}

\begin{document}

\preprint{L-edge-magnon}

\title{Effect of broken symmetry on resonant inelastic x-ray scattering 
from undoped cuprates}


\author{Jun-ichi Igarashi}%
\affiliation{%
Faculty of Science, Ibaraki University, Mito, Ibaraki 310-8512, Japan}

\author{Tatsuya Nagao}
\affiliation{%
Faculty of Engineering, Gunma University, Kiryu, Gunma 376-8515, Japan}

\date{\today}

\begin{abstract}
We study the magnetic excitation spectra of
resonant inelastic x-ray scattering (RIXS) 
at the $L$-edge from undoped cuprates
beyond the fast collision approximation.
We analyse the effect of the symmetry breaking ground state 
on the RIXS process of the Heisenberg model by using a projection procedure.
We derive the expressions of the scattering amplitude
in both one-magnon and two-magnon excitation channels.
Each of them consists of the isotropic and anisotropic contributions. 
The latter is a new finding and 
attributed to the long range order of the ground state.
The presence of anisotropic terms
is supported by numerical calculations on a two-dimensional spin cluster.
We express the RIXS spectra in the form of spin-correlation functions
with the coefficients evaluated on the cluster, and calculate the function 
in a two dimensional system within the $1/S$ expansion. 
Due to the anisotropic terms, the spectral intensities are considerably 
enhanced around momentum transfer $\textbf{q}=0$ in both one-magnon and 
two-magnon excitation channels. This finding may be experimentally 
confirmed by examining carefully the $\textbf{q}$-dependence of the spectra.
\end{abstract}

\maketitle

\section{Introduction\label{sect.1}}
Resonant inelastic x-ray scattering (RIXS) has attracted
much interest as a useful tool to investigate excited 
states in solids\cite{Ament2011-rmp}. 
The L-edge RIXS experiments have been 
recently carried out with high energy resolution in 
transition-metal compounds, which have revealed magnetic excitations 
as spectral peaks in the low-energy 
region\cite{Braicovich09,Braicovich10,Guarise10}. 
Starting from the undoped cuprates,
the activity spreads over, rapidly and widely, 
doped high-T$_{\rm{c}}$ cuprates\cite{Tacon2011,Dean2012,Dean2015},
nickelates\cite{Ghiringhelli2009}, pnictides\cite{Zhou2013},
$5d$ transition metal compounds\cite{J.Kim2012,J.Kim2012.327}, and so on.
Among them, the investigation on cuprates is one of the most active
fields due to the relation with high-T$_{\rm{c}}$ superconductivity.
Stimulated by these experiments, theoretical efforts
to elucidate mechanism of the magnetic RIXS in cuprates also 
have developed\cite{Igarashi2012-1,Igarashi2012-2,Kourtis2012,Nagao2012,
Chen2013,Jia2014,Benjamin.pp}. 
But rich information involved in the $L$-edge RIXS data such
as momentum and energy transfer dependence as well as polarization
dependence never ceases to require further reliable and convincing 
theories more than ever.

The $L$-edge resonance 
in undoped cuprates is described by the second-order dipole allowed 
process that
a $2p$-core electron is prompted to an empty $x^2-y^2$ orbital 
by absorbing photon and then an occupied $3d$ electron
combines with the core hole by emitting photon.
When the $3d$ orbital in the photo-emitting process is different from the one
in the photo-absorbing process, the excitations within the $3d$ orbitals
are brought about, which are called as d-d excitation
\cite{Ghiringhelli04}. When the $3d$ orbitals in the photo-absorbing and 
photo-emitting processes are the same $x^2-y^2$ orbital but their spins are 
different, magnetic excitations with spin flip could be 
generated\cite{Ament09,Haverkort2010}. Even if the spins are the same, 
the spin-conserving excitations could be brought about by 
the presence of the core hole during its finite lifetime
\cite{Igarashi2012-1,Igarashi2012-2,Kourtis2012}.
This process could be described only when it is treated beyond the fast 
collision approximation (FCA) that no relaxation could take place in the
intermediate state because of the short lifetime
\cite{Ament09,Haverkort2010}. 

In our previous papers\cite{Igarashi2012-1,Igarashi2012-2}, 
we have analysed the process 
leading to the final states in the second-order process,
and have clarified how the spin excitations are taken place around
the core hole site beyond the FCA. 
In one dimension, the analysis has been straightforward, 
since the spherical symmetry in spin space 
remains intact in the ground state, while 
in two dimensions under the antiferromagnetic ordered phase, 
the analysis has been rather complicated due to the breaking of spherical 
symmetry. In both cases, we have obtained the scattering amplitudes 
in an invariant form with respect to the polarization vectors
of the incident and scattered x-rays, and spin operators.
Disregarding possible effects of the symmetry breaking
ground state, we have
obtained spin excitations extending to neighbours of the core hole site. 
Such excitations have been clearly observed in a one-dimensional system
CaCu$_2$O$_3$\cite{Bisogni2014} and in two-dimensional systems 
Sr$_2$CuO$_2$Cl$_2$\cite{Guarise10} and La$_2$CuO$_4$\cite{Bisogni2012}.

However, under the presence of the antiferromagnetic long-range order, 
it may be reasonable to presume
that the scattering amplitudes include anisotropic terms
associated with the direction of the staggered moment, 
since the second-order process could
be affected by the anisotropy originated from the broken symmetry
of the ground state. 
This observation 
contrasts to neutron scattering, in which the scattering amplitude is 
directly described by the interaction Hamiltonian between the 
spins of neutron and electron.
The purpose of this paper is to clarify the presence of anisotropic terms
in the scattering amplitude under the presence of the spin long range order
by analysing the second-order process on a model of undoped cuprates, 
where the low-lying excitations are described by the Heisenberg model.
In the scattering amplitudes summarised in an invariant form,
we obtain the anisotropic terms, 
which include a vector characterizing the staggered moment. 

To estimate quantitative impact of the anisotropic terms, we evaluate them
by carrying out numerical analysis on spin clusters.
For a cluster of 13 spins, which is regarded as a model of two-dimensional
cuprate, various terms in the scattering amplitudes are calculated.
We verify the anisotropic terms have finite contributions.
If the connection of the anisotropic terms with the symmetry breaking
of the ground state is intrinsic,
the weights of anisotropic terms are expected to increase with
increasing antiferromagnetic long-range order parameter.
This anticipation is confirmed by the numerical calculation 
on a ring of 12 spins with varying the external staggered magnetic field,
which is given in Appendix \ref{App.C}.

Collecting up such amplitudes from all the Cu sites, 
we derive the RIXS spectra represented by spin correlation functions.
When we investigate the correlation functions, analysis on
a larger system might be preferable. This is because
the spin excitations propagate through the entire
crystal in the final state. Thus, we employ 
the $1/S$ expansion to the spin 
operators\cite{Holstein40}, which practically enables us to treat
an infinite system. As a result, 
we can express the RIXS spectra in terms of the correlation functions of
one-magnon and two-magnon contributions.
Since two magnons are excited close to each other, their mutual 
interaction is important. We treat multiple scattering of 
two magnons by following the 
method previously developed\cite{Igarashi2012-1}.
It turns out that the correlation functions for both
one-magnon and two-magnon channels
have anisotropic contributions in addition to isotropic ones.
We find that the anisotropic terms 
produce substantial enhancement on
the RIXS intensities
for momentum transfer $\textbf{q}$ close to the $\Gamma$ point
in both channels.
This shows a sharp contrast to the fact 
that the contributions from the isotropic terms 
vanish at $\textbf{q}=0$ in both channels. 
Experimentally, polarization analysis 
may help to clarify the existence of the anisotropic terms,
since the polarization dependence is completely different between the
one-magnon and the two-magnon spectra.

The present paper is organized as follows.
In section \ref{sect.2}, we describe the second-order dipole allowed 
process responsible for RIXS process. 
In section \ref{sect.3}, we analyse the RIXS process
paying attention to the influence of the symmetry breaking on the scattering
amplitude, in which anisotropic terms are derived in an invariant form.
In section \ref{sect.4}, we evaluate numerically the amplitudes
of creating excitations on a finite size of two-dimensional cluster 
under a molecular field on the boundary.
In section \ref{sect.5}, we derive
the RIXS spectra in terms of the spin-correlation functions, 
which are treated with the $1/S$ expansion to spin 
operators. The RIXS spectra consisting of one-magnon and two-magnon excitations 
are calculated. Section \ref{sect.6} is devoted to the concluding remarks.
In Appendix \ref{App.A}, absorption 
coefficients at the $L_{2}-$ and $L_{3}$-edges are briefly discussed.
A short comment on the projection procedure is given 
in Appendix \ref{App.B}.
In Appendix \ref{App.C}, we show how the expansion coefficients develop
with increasing the staggered moment in a finite-size ring
under the external staggered field. 
Appendix \ref{App.D} outlines the $1/S$ expansion 
in the Heisenberg model.

\begin{figure}
\includegraphics[width=8.0cm]{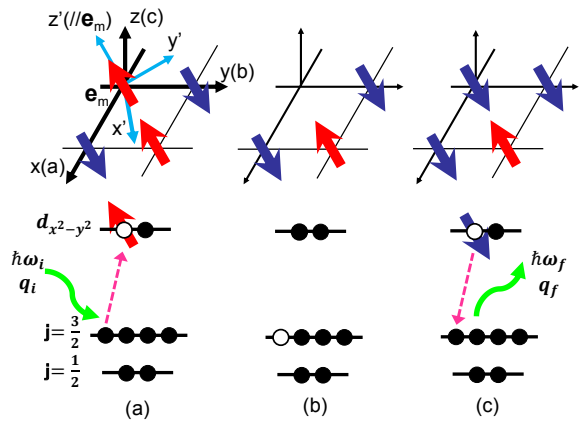}
\caption{\label{fig.RIXS}
A schematic illustration of the RIXS process at the $L_3$-edge of cuprates.
Filled and open circles stand for the states occupied by electrons and holes, 
respectively.
Red and blue arrows denote the magnetic moments of $3d$ holes.
Wavy green and pink dotted arrows represent the photon and
the transitions, respectively.
(a) An incident photon excites a core $2p_{\frac{3}{2}}$ electron
into the empty $3d$ state. The site of the excited electron is 
chosen as the origin of the crystal-fixed coordinate system with 
$x$, $y$, and $z$ axes (or also called as $a$, $b$, and $c$ axes).
The direction of the unit vector of the staggered magnetic moment
at the origin is called as $\textbf{e}_m$, which defines the
spin coordinate system with $x'$, $y'$, and $z'$ axes.
The spin quantization axis is coincident with $z'$ axis.
(b) In the intermediate state, spin degree of freedom vanishes
at the core-hole site.
(c) The excited electron recombines with the core-hole
by emitting the scattered photon.
}
\end{figure}

\section{Second-order dipole allowed process\label{sect.2}}
We briefly explain $L$-edge RIXS in cuprates (See figure \ref{fig.RIXS}). 
The RIXS process at the copper $L$-edge
may be described by the electric dipole ($E$1) transition between
the $2p$-core states and the $3d$ states. 
The $2p$ states are split into two levels with the total angular momentum 
$j=\frac{1}{2}$ and $\frac{3}{2}$, 
which are discriminated as $L_2$ and $L_3$ edges,
respectively, due to the strong spin-orbit interaction.
Because each Cu atom has one hole in the $x^2-y^2$ 
orbital in undoped cuprates such as La$_2$CuO$_4$ and
Sr$_2$CuCl$_2$O$_2$, we employ a hole picture.
Then, the $E_1$ transition may be expressed by the interaction 
between photon and hole as,
\begin{equation}
H_{\rm int}=w\sum_{\textbf{q}, \alpha,i, m, \sigma}
\frac{D^{\alpha}(j m,\sigma)}{\sqrt{2\omega_{\textbf{q}}}}
h_{i,j m}^{\dagger}c_{\textbf{q}\alpha}d_{i\sigma}
{\rm e}^{\textrm{i}\textbf{q}\cdot\textbf{r}_{i}} +{\rm H.c.},
\label{eq.tran1}
\end{equation}
where $c_{\textbf{q}\alpha}$ annihilates a photon 
with four-vector $q\equiv(\textbf{q},\omega_{\textbf{q}})$ and polarization $\alpha$.
The $h_{i,j m}^{\dagger}$ represents the creation operator of the $2p$ 
hole with $j m$ at site $i$, 
and $d_{i\sigma}$ denotes the annihilation operator of 
the $3d$ hole with the $x^2-y^2$ orbital and spin $\sigma$ at site $i$.
The $w$ is a constant proportional to 
$\int_0^{\infty}r^3R_{3d}(r)R_{2p}(r){\rm d}r$
where $R_{3d}(r)$ and $R_{2p}(r)$ are the 
radial wave-functions for 
the $3d$ and $2p$ states of Cu atom. 
The $D^{\alpha}(j m,\sigma)$ describes
the dependence of the $E$1 transition amplitude on the $2p$ core-hole
angular momentum and the spin of the $3d$ hole.

In the $E$1 transition at the $L_2$- and $L_3$-edges,
the initial photon having $q=q_{\rm i}$, $\alpha=\alpha_{\rm i}$
excites $2p$ core hole into empty $3d$ state,
which decays back into the $2p$ state by emitting the final
photon having $q=q_{\rm f}$, $\alpha=\alpha_{\rm f}$.
The RIXS spectra associated with this process may be expressed as 
\begin{eqnarray}
 W(q_{\rm{f}}\alpha_{\rm{f}};q_{\rm{i}}\alpha_{\rm{i}})
 &=& 2\pi\sum_{f'}\left|\sum_{n}
  \frac{\langle \Phi_{f'}|H_{\rm int}|n\rangle
        \langle n|H_{\rm int}|\Phi_{\rm{i}}\rangle}
       {E_{\rm{g}}+\omega_{\rm{i}}-E_n} \right|^2 \nonumber\\
 &\times&
\delta(E_{\rm{g}}+\omega_{\rm{i}}-E_{f'}
-\omega_{\rm{f}}),
\label{eq.optical}
\end{eqnarray} 
with $q_{\rm{i}}\equiv(\textbf{q}_{\rm{i}},\omega_{\rm{i}})$, 
$q_{\rm{f}}\equiv(\textrm{q}_{\rm{f}},\omega_{\rm{f}})$, 
$|\Phi_{\rm{i}}\rangle = c_{q_{\rm{i}}\alpha_{\rm{i}}}^{\dagger}
|\rm{g}\rangle|\rm{vac}\rangle$,
$|\Phi_{f'}\rangle=c_{q_{\rm{f}} \alpha_{\rm{f}}}^{\dagger}
|f'\rangle|\textrm{vac}\rangle$,
where $|\rm{g}\rangle$ and $|f'\rangle$ represent the ground state and
excited states of the matter with energy $E_{\rm{g}}$ and $E_{f'}$, 
respectively. 
Note that ${\rm f}$ refers to the final state of the photon and $f'$ refers to
the excited state of the electron.
The $|\rm{vac}\rangle$ is the vacuum state for photons.
The eigenstate and its energy of the intermediate
state are referred to as $|n \rangle$ and $E_n$, respectively.

Incidentally, since the final state in the absorption coefficient $A(\omega)$ 
is the intermediate state in the RIXS, we have
\begin{equation}
 A(\omega)=2\pi\sum_{n}|\langle n|H_{\rm int}|\Phi_{\rm{i}}\rangle|^2
           \delta(E_{\rm{g}}+\omega-E_n).
\end{equation}
The explicit form is summarised in Appendix \ref{App.A}.

\section{\label{sect.3}Magnetic excitations around the core-hole site}
In undoped cuprates, 
the low-energy excitations may be well described by
the two-dimensional antiferromagnetic Heisenberg Hamiltonian
on a square lattice,
\begin{equation}
 H_{\rm mag}=J \sum_{\langle i,i'\rangle} 
                   \textrm{S}_i\cdot \textrm{S}_{i'},
\end{equation}
where $\textrm{S}_{i}$ denotes the spin one half operator at site $i$,
and $\langle i,i'\rangle$ indicates that the summation runs over 
nearest-neighbour pairs. 
Since our focus is not on a discussion of the magnetic
dispersion, we have adopted the exchange interaction $J$
only between the nearest neighbour sites. 
In the thermodynamic limit, the ground state of $H_{\rm mag}$ on a
square lattice is spontaneous symmetry broken phase, that is, 
long-range ordering antiferromagnetic phase.

We write the ground state $|g\rangle$ of $H_{\rm mag}$ as
\begin{equation}
 |g\rangle = |\uparrow\rangle|\psi_0^{\uparrow}\rangle
           + |\downarrow\rangle|\psi_0^{\downarrow}\rangle,
\end{equation}
where $|\uparrow\rangle$ and $|\downarrow\rangle$ represent the spin
states at the origin, and $|\psi_0^{\uparrow}\rangle$ and 
$|\psi_0^{\downarrow}\rangle$ are constructed 
by the bases of the rest of spins.
We assume that a core hole is created at the origin as a result
of absorbing photon (figure \ref{fig.RIXS}). 
In the intermediate state, the spin degrees of freedom is lost at the 
core-hole site, since the $3d$ hole in the $x^2-y^2$ orbital is
annihilated by the $2p$-$3d$ dipole transition. 
Note that the Hamiltonian 
in the intermediate state is similar to that for a system
with a non-magnetic impurity introduced into 
antiferromagnet\cite{Tonegawa68,Wan93}.

Denoting $|\phi_{\eta}\rangle$ as the eigenstate of the 
intermediate Hamiltonian with eigenvalue $\epsilon'_{\eta}$,
we can express the second-order amplitude in (\ref{eq.optical}) 
as
\begin{eqnarray}
\sum_n \frac{H_{\rm int}|n\rangle \langle n|H_{\rm int}|\rm{g}\rangle}
            {\omega_{\rm{i}}+E_{\rm{g}}-E_n}
   &\propto& \sum_{m, \sigma, \sigma'} 
D^{\alpha_{\rm{f}}}(j m,\sigma)^{*}
                             D^{\alpha_{\rm{i}}}(j m,\sigma') \nonumber \\
&\times&
   \sum_{\eta}|\sigma\rangle |\phi_{\eta}\rangle R(\epsilon'_{\eta})
   \langle\phi_{\eta}|\psi_{0}^{\sigma'}\rangle,
\label{eq.process1}
\end{eqnarray}
with
\begin{equation}
 R(\epsilon'_{\eta}) =
 \frac{1}{\omega_{\rm{i}}+\epsilon_{\rm{g}} 
-\epsilon_{\rm{core}}+i\Gamma - \epsilon'_{\eta}},
\end{equation}
where $\epsilon_{\rm{g}}$ represents the ground state 
energy of $H_{\rm mag}$.
The $\epsilon_{\rm{core}}$ denotes 
the energy required to create the $3d^{10}$-configuration and
a $2p$ core hole in the state $|j m\rangle$.
The $\Gamma$ stands for the life-time broadening width of the core hole;
$\Gamma\sim 0.3$ eV at the Cu $L_3$ edge. 
Notice that the scattering amplitude (6) and those investigated in 
the remaining of this section are originated from the excitation of the 
single electron at the origin. A whole scattering intensity will be given 
by collecting up the amplitudes from all Cu sites.

In the scattering amplitudes leading to those excited states,
we seek the invariant form with the polarization
vectors $\mbox{\boldmath{$\alpha$}}_{\rm{i}}$ and 
$\mbox{\boldmath{$\alpha$}}_{\rm{f}}$ 
of the incident and scattered x-rays, spin operators $\textrm{S}_i$, and the
unit vector of the staggered moment $\textrm{e}_{\rm{m}}$. 
To this end, it is convenient to consider a general situation that 
$\textrm{e}_{\rm{m}}$ is pointing to an arbitrary direction, 
which is denoted as axis $z'$. 
Then, for spin operators of the $3d$ electron, 
coordinate frame of $x'$, $y'$, $z'$ axes is prepared 
(See figure \ref{fig.RIXS} (a)).
On the other hand, the $3d$ orbitals as well as 
spin and orbital of the core hole are described in 
the crystal-fixed coordinate frame with $x$, $y$, and $z$ axes.
Since the definition of the spin coordinate system
and that of the crystal-fixed system are independent, 
we can relate them by any method which can describe the transformation 
from the one to the other.  We adopt here the rotation of 
the Euler angles $\alpha$, $\beta$, and $\gamma$ as 
the transformation from the crystal-fixed to the spin coordinate system
\cite{Rose1957}. Our final formulae do not depend on the specific choice
of the Euler angles.

The $D^{\mu}(j m,\sigma)$ in this definition is given 
in table I of \cite{Igarashi2012-1}. Then we introduce 
$P_{\sigma}^{(0)}(j;\alpha_{\rm{f}},\alpha_{\rm{i}})$
and $P_{\sigma}^{(1)}(j;\alpha_{\rm{f}},\alpha_{\rm{i}})$ by
\begin{eqnarray}
\sum_{m}D^{\alpha_{\rm{f}}}(j m,\sigma)^{*}
D^{\alpha_{\rm{i}}}(j m,\sigma)
&\equiv& P_{\sigma}^{(0)}(j ;\alpha_{\rm{f}},\alpha_{\rm{i}}), \\
\sum_{m}D^{\alpha_{\rm{f}}}(j m,\sigma)^{*}
D^{\alpha_{\rm{i}}}(j m,-\sigma) 
&\equiv& P_{\sigma}^{(1)}(j ;\alpha_{\rm{f}},\alpha_{\rm{i}}),
\end{eqnarray}
where $-\sigma$ represent $\downarrow$ and $\uparrow$
for $\sigma=\uparrow$ and $\downarrow$, respectively.
The $P_{\sigma}^{(0)}(j ;\alpha_{\rm{f}},\alpha_{\rm{i}})$ and 
$P_{\sigma}^{(1)}(j ;\alpha_{\rm{f}},\alpha_{\rm{i}})$ correspond to 
the spin-conserving and the spin-flipping processes, respectively.
Polarizations of x-ray are along the $x$, $y$, 
and $z$ axes defined in the original crystal axes.
Since the following analysis is confined to the $L_3$-edge,
we fix $j =\frac{3}{2}$ and omit the argument 
in the expressions of
$P_{\sigma}^{(0)}(j ;\alpha_{\rm{f}},\alpha_{\rm{i}})$ and 
$P_{\sigma}^{(1)}(j ;\alpha_{\rm{f}},\alpha_{\rm{i}})$.
We list all the non-zero values of them for $j =\frac{3}{2}$ 
below.
\begin{eqnarray}
P_{\sigma}^{(0)}(x,x)&=&P_{\sigma}^{(0)}(y,y)
=\frac{2}{15}, \label{eq.P0.d}\\
P_{\sigma}^{(0)}(x,y)&=&-P_{\sigma}^{(0)}(y,x)
=-\rm{sgn}(\sigma) \frac{\rm{i}}{15}\cos\beta, \label{eq.P0.od} \\
P_{\sigma}^{(1)}(x,y)&=&-P_{\sigma}^{(1)}(y,x)
=\frac{\rm{i}}{15}\exp[\rm{i} \gamma \rm{sgn}(\sigma)] \sin\beta, 
\nonumber\\
\label{eq.P1.od}
\end{eqnarray}
where $\rm{sgn}(\sigma)$ gives $+1$ and $-1$ for $\sigma=\uparrow$
and $\downarrow$, respectively.
Note that $P^{(0)}(\alpha_{\rm{f}},\alpha_{\rm{i}})$ and 
$P_{\sigma}^{(1)}(\alpha_{\rm{f}},\alpha_{\rm{i}})$ are zero if
$\alpha_{\rm{i}}=z$ and/or $\alpha_{\rm{f}}=z$. 
This results from the fact that 
the process is restricted with the hole of the $x^2-y^2$ orbital in 
the ground state.

\subsection{Scattering channel with changing polarization}
As seen from (\ref{eq.P0.od}) and (\ref{eq.P1.od}),
both $P_{\sigma}^{(0)}(\alpha_{\rm{f}},\alpha_{\rm{i}})$ and 
$P_{\sigma}^{(1)}(\alpha_{\rm{f}},\alpha_{\rm{i}})$ have off-diagonal 
elements with $\alpha_{\rm{i}}$ and $\alpha_{\rm{f}}$. 
This implies that the scattering channel with changing photon polarization
includes both the spin-flipping and spin-conserving processes.
Let us investigate them separately in the following.

\subsubsection{Spin-flipping process}
The final state arising from the spin-flipping process may be written as
\begin{equation}
  |F\rangle \equiv
\sum_{\sigma}P_{\sigma}^{(1)}(\alpha_{\rm{f}},\alpha_{\rm{i}})
  |\sigma\rangle\sum_{\eta}|\phi_{\eta}\rangle R(\epsilon'_{\eta})
  \langle\phi_{\eta}|\psi_{0}^{-\sigma}\rangle.
\label{eq.spin-flip}
\end{equation}
Assuming the magnetic excitation associated with the creation
of core-hole at site 0 has a local character around the core-hole site,
we approximate $|F\rangle$ by a linear combination of the states 
$|\psi_{1}^{(\pm)}\rangle=S_{0}^{\pm}|\rm{g}\rangle$ and
$|\psi_{2}^{(\pm)}\rangle=X^{\pm}|\rm{g}\rangle$,
where $\textrm{X}=\frac{1}{z}\sum_{j}\textrm{S}_j$ with $j$ running over the
nearest neighbour sites around the core-hole site.
The number of the nearest neighbour sites $z$ is four and two for 
two and one dimensions, respectively.
Spin raising and lowering operators on the core-hole site 
and neighbouring site are defined as
$S_{0}^{\pm}=S_{0}^{x'}\pm \textrm{i} S_{0}^{y'}$ 
and $X^{\pm}=X^{x'}\pm \textrm{i} X^{y'}$, respectively. 

Since the $|\psi_n^{(\pm)}\rangle$'s are not orthogonal to each 
other nor normalized,
we need to introduce the density matrices $(\hat{\rho}^{(\pm)})_{i,j}
=\langle\psi_{i}^{(\pm)}|\psi_{j}^{(\pm)}\rangle$ 
to project $|F\rangle$ onto $|\psi_i^{(\pm)}\rangle$'s. 
A procedure to determine
the expansion coefficients is given in Appendix \ref{App.B}
where the projection formalism is utilised.
It may seem strange the non-orthonormal set is used in the expansion.
However, since the procedure described in Appendix \ref{App.B} can determine
the expansion coefficients uniquely for the finite number of the 
projected states, the non-orthonormal set can have a one-to-one 
correspondence with some orthonormal set, for instance, 
by means of Gram-Schmidt process.
Since the physical meaning of each
element of the non-orthonormal set is much clearer than that of the orthonormal 
one, we use the former.

Then, the final state is approximately expressed as
\begin{eqnarray}
  |F\rangle &\approx& 
     \sum_{\sigma} P_{\sigma}^{(1)}(\alpha_{\rm{f}},\alpha_{\rm{i}})
     \sum_{ij,\nu=\pm}|\psi_i^{(\nu)}\rangle
     (\hat{\rho}^{(\nu)-1})_{i,j} \langle\psi_{j}^{(\nu)}| \nonumber \\
&\times& 
\sum_{\eta} |\sigma\rangle 
|\phi_{\eta}\rangle R(\epsilon'_{\eta})
     \langle\phi_{\eta}|\psi_{0}^{-\sigma}\rangle.
\label{eq.proj1}
\end{eqnarray}
This expression is rearranged as
\begin{eqnarray}
 |F\rangle &\approx& 
P_{\downarrow}^{(1)}(\alpha_{\rm{f}},\alpha_{\rm{i}})
 \left[f_{0,\downarrow}^{(1)}(\omega_i)S_0^{-}|\rm{g}\rangle
      +f_{1,\downarrow}^{(1)}(\omega_i)X^{-}|\rm{g}\rangle \right]
\nonumber\\
    &+& P_{\uparrow}^{(1)}(\alpha_{\rm{f}},\alpha_{\rm{i}})
 \left[f_{0,\uparrow}^{(1)}(\omega_i)S_0^{+}|\rm{g}\rangle
      +f_{1,\uparrow}^{(1)}(\omega_i)X^{+}|\rm{g}\rangle \right]. 
\nonumber\\
\label{eq.spin-flip1}
\end{eqnarray}
The coefficients are given by 
\begin{eqnarray}
 f_{0,\downarrow}^{(1)}(\omega_i) &=&
    \sum_{j=1,2} (\hat{\rho}^{(-)-1})_{1,j}
   \nonumber \\
&\times& \langle\psi_{j}^{(-)}|
\sum_{\eta} |\downarrow\rangle 
|\phi_{\eta}\rangle R(\epsilon'_{\eta})
     \langle\phi_{\eta}|\psi_{0}^{\uparrow}\rangle, \label{eq.f0d} \\
 f_{0,\uparrow}^{(1)}(\omega_i) &=&
    \sum_{j=1,2} (\hat{\rho}^{(+)-1})_{1,j}
  \nonumber\\
&\times& \langle\psi_{j}^{(+)}| \sum_{\eta} |\uparrow\rangle 
|\phi_{\eta}\rangle R(\epsilon'_{\eta})
     \langle\phi_{\eta}|\psi_{0}^{\downarrow}\rangle. \label{eq.f0u}
\end{eqnarray}
The $f_{1,\sigma}^{(1)}(\omega_i)$ can be constructed
from $f_{0,\sigma}^{(1)}(\omega_i)$ by replacing 
$(\hat{\rho}^{(\pm)-1})_{1,j}$ with $(\hat{\rho}^{(\pm)-1})_{2,j}$.
Let us examine each coefficients appeared in (\ref{eq.spin-flip1}).
We suppose that the core hole site belongs to `up spin' sublattice.
This does not mean $S_0^+ |g\rangle =0$ when $|g\rangle$ is the 
symmetry broken antiferromagnetic 
ground state. That is, the spin can be raised 
even at the `up spin site'. Then, for example, if spin-flip excitation takes
place at the core-hole site, two channels, from up spin to down spin 
and vice versa should be survived. Each cahnnel experiences different 
surroundings in the intermediate state through the second order process, 
which is materialized due to the fact that the core-hole has a finite 
life-time.
As a result, both channels aquire different values of the coefficients.
Similar explanation is also valid for the spin-flip process
at the nearest neighbour sites.

In the presence of the antiferromagnetic long-range order,
the coefficients $f_{0,\sigma}^{(1)}(\omega_i)$ and 
$f_{1,\sigma}^{(1)}(\omega_i)$ for $\sigma=\uparrow$ are expected to be 
different from those for $\sigma=\downarrow$.
Then, let us divide them into two parts as follows.
\begin{eqnarray}
   f_{0,\sigma}^{(1)}(\omega_i) &=& f_{0}^{(1)}(\omega_i)
  + \rm{sgn}(\sigma) \Delta_{\perp,0}^{(1)}(\omega_i), 
\label{eq.coe1}\\
   f_{1,\sigma}^{(1)}(\omega_i) &=& f_{1}^{(1)}(\omega_i)
  + \rm{sgn}(\sigma) \Delta_{\perp,1}^{(1)}(\omega_i). 
\label{eq.coe2}
\end{eqnarray}
It has been confirmed that $\Delta_{\perp,0}(\omega_i)$ $=$ 
$\Delta_{\perp,1}(\omega_i)$ $=0$ and 
$f_{n,\uparrow}^{(1)}(\omega_i)=f_{n,\downarrow}^{(1)}(\omega_i)$
in the absence of the
long-range order for one-dimensional system \cite{Igarashi2012-2}. 
Therefore, the $\Delta_{\perp,n}^{(1)}(\omega_i)$ 
stands for the \textit{anisotropic} part of the coefficient,
while $f_n^{(1)}(\omega_i)$ represents the \textit{isotropic}
part of coefficient.
Note that the anisotropic coefficient $\Delta_{\perp,0}(\omega_i)$ and 
the coefficients for the excitations on neighbouring sites
$f_{1,\sigma}^{(1)}(\omega_i)$ would not come out in the FCA, 
since the relaxation process in the intermediate state is disregarded.
Inserting (\ref{eq.coe1}) and (\ref{eq.coe2}) into 
(\ref{eq.spin-flip1}) with the help of (\ref{eq.P1.od}), 
we notice that (\ref{eq.spin-flip1}) with the Euler angles
$\alpha$, $\beta$, $\gamma$ constitute an invariant form
(see (3.18) in \cite{Igarashi2012-1} for isotropic terms).
The result is given by
\begin{eqnarray}
  |F\rangle &\approx& \frac{2 \rm{i}}{15} 
(\mbox{\boldmath{$\alpha$}}_{\rm{i}\perp}\times 
  \mbox{\boldmath{$\alpha$}}_{\rm{f}\perp}) \cdot \Bigl\{ 
       f_{0}^{(1)}(\omega_{\rm{i}})\textrm{S}_{0\perp}
      +f_{1}^{(1)}(\omega_{\rm{i}})\textrm{X}_{\perp}
\nonumber\\
  &- & \rm{i} \left[
 \Delta_{\perp,0}^{(1)}(\omega_{\rm{i}})\textrm{e}_{\rm{m}}
\times \textrm{S}_{0\perp}
+\Delta_{\perp,1}^{(1)}(\omega_{\rm{i}})\textrm{e}_{\rm{m}}
\times \textrm{X}_{\perp}\right]
\Bigr\}|g\rangle,\nonumber\\
\label{eq.spin-flip}
\end{eqnarray}
where $\mbox{\boldmath{$\alpha$}}_{\rm{i}\perp}$ and 
$\mbox{\boldmath{$\alpha$}}_{\rm{f}\perp}$, respectively,
are polarization vectors of the incident and scattered photon,
which are projected onto the $a$-$b$ plane.
Operators $\textrm{S}_{0\perp}$ and $\textrm{X}_{\perp}$, respectively, are 
$\textrm{S}_{0}$ and $\textrm{X}$, which are
projected onto the plane perpendicular to the direction of the 
staggered magnetic moment. 

\subsubsection{Spin-conserving process}
According to (\ref{eq.process1}), the spin-conserving process may be
written as
\begin{equation}
 |F' \rangle \equiv
\sum_{\sigma} P_{\sigma}^{(0)}(\alpha_{\rm{f}},\alpha_{\rm{i}})
\sum_{\eta} |\sigma\rangle
|\phi_{\eta}\rangle R(\epsilon'_{\eta})
  \langle\phi_{\eta}|\psi_{0}^{\sigma}\rangle ,
\end{equation}
where the off-diagonal elements with the polarizations are used for
$P_{\sigma}^{(0)}(\alpha_{\rm{f}},\alpha_{\rm{i}})$.
We approximate $|F' \rangle$ by a linear combination of
the states $|\psi_1\rangle=|\rm{g}\rangle$,
$|\psi_2\rangle=S_0^{z'}|\rm{g}\rangle$, and 
$|\psi_3\rangle=X^{z'}|\rm{g}\rangle$.
Since these states are not orthogonal to each other nor normalized,
we repeat the analysis that utilises the density matrix
$\hat{\rho}_{i,j} \equiv \langle \psi_i | \psi_j \rangle$.
Hence the final state in this channel is approximately expressed as
\begin{eqnarray}
  |F' \rangle &\approx& 
     \sum_{\sigma} P_{\sigma}^{(0)}(\alpha_{\rm{f}},\alpha_{\rm{i}})
     \sum_{ij}|\psi_i\rangle
     (\hat{\rho}^{-1})_{i,j}\langle\psi_{j} | \nonumber\\
  &\times&   
 \sum_{\eta}|\sigma\rangle |\phi_{\eta}\rangle R(\epsilon'_{\eta})
     \langle\phi_{\eta}|\psi_{0}^{\sigma}\rangle .
\label{eq.proj2}
\end{eqnarray}
This relation is rewritten as 
\begin{eqnarray}
 |F' \rangle &\approx& 
\frac{2 \rm{i}}{15} (\mbox{\boldmath{$\alpha$}}_{\rm{i}\perp}\times 
  \mbox{\boldmath{$\alpha$}}_{\rm{f}\perp}) 
  \cdot
  \left[g_{0}^{(1)}(\omega_{\rm{i}})\textrm{S}_{0\parallel}
       +g_{1}^{(1)}(\omega_{\rm{i}})\textrm{X}_{\parallel}\right]
|\rm{g}\rangle, \nonumber\\
\label{eq.spin-conserv}
\end{eqnarray}
where $\textrm{S}_{0\parallel}$ and $\textrm{X}_{\parallel}$, respectively,
represent the vector operators of $\textrm{S}_0$ and $\textrm{X}$ 
parallel to the direction of the staggered magnetic moment.
Note that the amplitude associated with $|\psi_1 \rangle$ is omitted.
The definition of the expansion coefficient $g_n^{(1)}(\omega_{\rm i})$ 
is inferred from the projection procedure in Appendix \ref{App.B}.
We have already confirmed
that $g_0^{(1)}(\omega_{\rm{i}})$ and 
$g_1^{(1)}(\omega_{\rm{i}})$ were equivalent to $f_0^{(1)}(\omega_{\rm{i}})$ 
and $f_1^{(1)}(\omega_{\rm{i}})$, respectively,
in the absence of long-range order \cite{Igarashi2012-2}.\cite{Com1}
Therefore, it is natural, in the presence of long-range order, 
to write them as 
\begin{eqnarray}
 g_{0}^{(1)}(\omega_{\rm{i}}) &=& f_0^{(1)}(\omega_{\rm{i}})
       -\textrm{i}\Delta_{\parallel,0}^{(1)}(\omega_{\rm{i}}), \\
 g_{1}^{(1)}(\omega_{\rm{i}}) &=& f_1^{(1)}(\omega_{\rm{i}})
       -\textrm{i}\Delta_{\parallel,1}^{(1)}(\omega_{\rm{i}}).
\end{eqnarray}
Here $\Delta_{\parallel,0}^{(1)}(\omega_{\rm{i}})$
and $\Delta_{\parallel,1}^{(1)}(\omega_{\rm{i}})$
correspond to the anisotropic contributions of the coefficients.

Combining the spin-conserving term (\ref{eq.spin-conserv}) to 
the spin-flipping term (\ref{eq.spin-flip}), we finally have
\begin{eqnarray}
&& \hspace*{-0.5cm} |F\rangle_1  \equiv |F\rangle +|F'\rangle
 \nonumber\\
&\approx& \frac{2 \rm{i}}{15} 
  (\mbox{\boldmath{$\alpha$}}_{\rm{i}\perp}\times 
  \mbox{\boldmath{$\alpha$}}_{\rm{f}\perp}) 
 \cdot 
 \left[f_{0}^{(1)}(\omega_{\rm{i}})\textrm{S}_{0}
-\textrm{i} \Delta_{\perp,0}^{(1)}(\omega_{\rm{i}})
\textrm{e}_{\rm{m}}\times \textrm{S}_{0}
       \right. \nonumber\\
&& \left. 
-\textrm{i}\Delta_{\parallel,0}^{(1)}(\omega_{\rm{i}})
\textrm{e}_{\rm{m}}(\textrm{e}_{\rm{m}} \cdot \textrm{S}_{0})
\right]| \rm{g} \rangle 
\nonumber\\
  &+& \frac{2 \textrm{i}}{15} 
  (\mbox{\boldmath{$\alpha$}}_{\rm{i}\perp}\times 
  \mbox{\boldmath{$\alpha$}}_{\rm{f}\perp})
\cdot \left[
   f_{1}^{(1)}(\omega_{\rm{i}}) \textrm{X}
-\textrm{i} \Delta_{\perp,1}^{(1)}(\omega_{\rm{i}}) \textrm{e}_{\rm{m}}
\times \textrm{X}
\right. \nonumber \\
&& \left. 
-\textrm{i} \Delta_{\parallel,1}^{(1)}(\omega_{\rm{i}})
\textrm{e}_{\rm{m}} (\textrm{e}_{\rm{m}}\cdot \textrm{X})
\right]|\textrm{g}\rangle.
\label{eq.pol-flip}
\end{eqnarray}
The terms containing $\textrm{e}_{\rm{m}}$ represent 
the effect of the long range order, that is, 
that of the broken symmetry in spin space.
If $\textrm{e}_{\rm{m}}$ is defined on the A sublattice and the same 
$\textrm{e}_{\rm{m}}$ is used on the B sublattice, 
$\Delta_{\parallel,0}^{(1)}(\omega_{\rm{i}})$ and 
$\Delta_{\parallel,1}^{(1)}(\omega_{\rm{i}})$, respectively,
take the same value in both sublattices.
On the other hand,  the values of 
$\Delta_{\perp,0}^{(1)}(\omega_{\rm{i}})$ and
$\Delta_{\perp,1}^{(1)}(\omega_{\rm{i}})$ in sublattice B, respectively, 
are obtained by changing entire sign of those in sublattice A, respectively.

\subsection{Scattering channel without changing polarization}
Since only $P_{\sigma}^{(0)}(\alpha_{\rm{f}},\alpha_{\rm{i}})$ has the 
non-zero diagonal elements with $\alpha_{\rm{i}}$
and $\alpha_{\rm{f}}$, (\ref{eq.process1}) may be expressed as 
\begin{equation}
 |F\rangle_2 \equiv
\sum_{\sigma} P_{\sigma}^{(0)}(\alpha,\alpha)
  |\sigma\rangle\sum_{\eta}|\phi_{\eta}\rangle R(\epsilon'_{\eta})
  \langle\phi_{\eta}|\psi_{0}^{\sigma}\rangle.
\end{equation}
We see that the FCA could not give rise to spin excitations 
in this process because 
the diagonal element $P_{\sigma}^{(0)}(\alpha,\alpha)$ 
is independent of $\sigma$. 
Since the total spin is conserved, 
$|F\rangle_2$ may be expressed by 
$|g\rangle$, $S_0^{z'}|\rm{g}\rangle$, $X^{z'}|\rm{g}\rangle$,
$S_0^{z'}X^{z'}|\rm{g}\rangle$, and 
$\frac{1}{2}(S_0^{+}X^{-}+S_0^{-}X^{+})|\rm{g}\rangle$. 
Similar to the procedure resorted in the
preceding subsection, $|F\rangle_{2}$ is approximated by a linear
combination of these states with the help of the density matrix. 
Hence $|F\rangle_{2}$ is approximately expressed as
\begin{eqnarray}
 |F\rangle_2 &\approx& \frac{2}{15}
  (\mbox{\boldmath{$\alpha$}}_{\rm{i}\perp}\cdot
  \mbox{\boldmath{$\alpha$}}_{\rm{f}\perp}) 
  f_2^{(2)}(\omega_{\rm{i}}) \textrm{S}_0\cdot \textrm{X}|\rm{g}\rangle \nonumber\\
 &+& \frac{2}{15}
  (\mbox{\boldmath{$\alpha$}}_{\rm{i}\perp}\cdot
  \mbox{\boldmath{$\alpha$}}_{\rm{f}\perp}) 
   \textrm[\Lambda^{(2)}(\omega_{\rm{i}})
    (\textrm{e}_{\rm{m}}\cdot \textrm{S}_0)(\textrm{e}_{\rm{m}}\cdot \textrm{X}) 
\nonumber\\
    &+&\Delta_{\parallel,0}^{(2)}(\omega_{\rm{i}})
(\textrm{e}_{\rm{m}}\cdot \textrm{S}_{0})
   +\Delta_{\parallel,1}^{(2)}(\omega_{\rm{i}})(\textrm{e}_{\rm{m}}\cdot
\textrm{X})\textrm]|\rm{g}\rangle,
\label{eq.pol-conserve}
\end{eqnarray}
where the amplitude associated with $|\rm{g}\rangle$ is omitted.
The terms containing $\textrm{e}_{\rm{m}}$ represent 
the effect of broken symmetry in spin space.
The expansion coefficients for 
$S_0^{z'}|\rm{g}\rangle$ and $X^{z'}|\rm{g}\rangle$ are
denoted as $\Delta_{\parallel,0}^{(2)}(\omega_{\rm{i}})$ and 
$\Delta_{\parallel,1}^{(2)}(\omega_{\rm{i}})$, respectively, 
while those defined for $S_0^{z'}X^{z'}|\rm{g}\rangle$ and 
$\frac{1}{2}(S_0^{+}X^{-}+S_0^{-}X^{+})|\rm{g}\rangle$ are
divided into the isotropic term $f_2^{(2)}(\omega_{\rm i})$
and anisotropic term $\Lambda^{(2)}(\omega_{\rm i})$.
If $\textrm{e}_{\rm{m}}$ is defined on the A sublattice and 
the same $\textrm{e}_{\rm{m}}$ is used on the B sublattice, 
$\Lambda^{(2)}(\omega_{\rm{i}})$ take the same value in both sublattices.
On the other hand,  the values of 
$\Delta_{\parallel,0}^{(2)}(\omega_{\rm{i}})$ and 
$\Delta_{\parallel,1}^{(2)}(\omega_{\rm{i}})$ in sublattice B, respectively, 
are obtained by changing entire sign of those in sublattice A, respectively.

\begin{table*}
\caption{\label{table.3}
Various coefficients in units of $1/J$ in the two-dimensional cluster of
13 spins: (a) coefficients for isotropic terms and (b) coefficients for
anisotropic terms. The incident photon energy $\omega_{\rm{i}}$ is set 
to give the maximum absorption coefficient.}
\footnotesize
\begin{tabular}{ccccc}
\hline
\hline
     & \multicolumn{2}{c}{(a) Isotropic coefficients} & 
       \multicolumn{2}{c}{(b) Anisotropic coefficients} \\
\hline
$\Gamma/J$   & $f_{0}^{(1)}(\omega_{\rm{i}})$ & 
               $f_2^{(2)}(\omega_{\rm{i}})$ & 
               $\Delta_{\perp,0}^{(1)}(\omega_{\rm{i}})$ &
               $\Lambda^{(2)}(\omega_{\rm{i}})$ \\ 
\hline
       $2.4$ & $(-0.087,-0.367)$ & $(0.177,-0.320)$ & $(-0.093,0.048)$ & $(0.071,-0.002)$ \\
       $1.0$ & $(-0.263,-0.656)$ & $(0.082,-0.891)$ & $(-0.209,0.334)$ & $(0.182,-0.207)$ \\
\hline
\hline
\end{tabular}
\end{table*}
\normalsize

\section{Evaluation of the coefficients\label{sect.4}}
\begin{figure}
\includegraphics{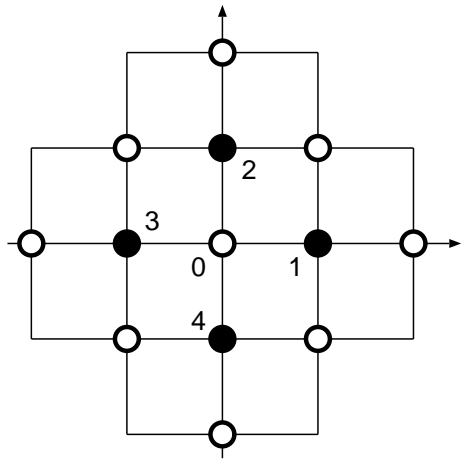}
\caption{\label{fig.square}
A cluster of 13 spins used to evaluate the coefficients.
The spin at site 0 is annihilated in the intermediate state.
Spins on the boundary are subjected to the molecular field 
from spins outside the cluster.
}
\end{figure}

Various coefficients defined in the preceding section could be evaluated by 
diagonalizing the Heisenberg Hamiltonian on finite-size clusters. 
Since the excitations are localized around the core-hole site, 
the calculation on small clusters may give reliable estimates 
to the coefficients.
We consider a cluster of 13 spins shown in figure \ref{fig.square}.
A complication is that analysis on finite-size cluster cannot provide
spontaneous symmetry breaking ground state.
In order to break the spherical symmetry in spin space, we assume that 
the spins on the boundary are subjected to the molecular field, 
$-J|\langle S_{0}^{z'}\rangle|$, per bond.
The expectation value of $S_{0}^{z'}$
is determined self-consistently as
$\langle S_{0}^{z'}\rangle=0.394$. 
The coefficients in the RIXS process are evaluated 
by diagonalizing the Hamiltonian matrices in the ground state and in the
intermediate state. Table \ref{table.3} shows the calculated results 
at $\omega_i$ giving the maximum absorption coefficient with $\Gamma/J=2.4$
and $1.0$.  The values for $\Gamma/J=2.4$ may correspond to La$_2$CuO$_4$ and 
Sr$_2$CuO$_2$Cl$_2$. 
Note that the coefficients have dimensions of (energy)$^{-1}$
as seen from right-hand side of (\ref{eq.process1}).
The coefficients not shown there are small, and will be neglected 
in the calculation of the RIXS spectra in the next section.

As seen from (\ref{eq.pol-flip}) and (\ref{eq.pol-conserve}),
it is obvious qualitatively that 
the origin of the anisotropic terms, which include the unit
vector representing the staggered moment ($\textrm{e}_{\rm{m}}$),
is attributed to the broken symmetry of the ground state in spin space.
In quantitative sense, the magnitudes of such terms are 
expected to develop as increasing the staggered moment. 
This is confirmed in Appendix  
\ref{App.C} for a finite-size ring of spins.

\section{Analysis of RIXS spectra from undoped cuprates\label{sect.5}}
Now, we are in a position to calculate the RIXS spectra.
It is preferable to treating a larger system
since the spin excitations propagate through the entire crystal
in the final state.
Thus, we employ the results of the 
$1/S$ expansion to the spin operators, which practically 
corresponds to taking into account of an infinite system effect
as well as the interaction among the magnetic excitations.
In doing so, we proceed the analysis by dividing the RIXS spectra
into two channels, with and without changing photon polarization.

\subsection{Scattering channel with changing polarization}

Since $\mbox{\boldmath{$\alpha$}}_{\rm{f}\perp}$ and 
$\mbox{\boldmath{$\alpha$}}_{\rm{i}\perp}$ are polarization 
vectors projected onto 
the $a$-$b$ plane, $\mbox{\boldmath{$\alpha$}}_{\rm{f}\perp}\times  
\mbox{\boldmath{$\alpha$}}_{\rm{i}\perp}$ is parallel to the $c$ axis.
In undoped cuprates such as La$_2$CuO$_4$ and Sr$_2$CuO$_2$Cl$_2$,
the staggered magnetization aligns along the $(1,1,0)$ 
direction in the CuO$_2$ plane\cite{Vaknin87}. 
Therefore the anisotropic terms proportional to $\textrm{e}_{\rm{m}}$ 
could not come out. 
We collect up the remaining amplitudes from all Cu sites, where
(\ref{eq.pol-flip}) is multiplied by the weight 
$\exp(\rm{i} \textrm{q}\cdot \textrm{r}_i)$ 
at the core-hole site $\textrm{r}_i$ with momentum transfer
$\textrm{q}\equiv \textrm{q}_{\rm{i}}-\textrm{q}_{\rm{f}}$.
Thereby we obtain
\begin{eqnarray}
 &&W(q_{\rm{f}},\mbox{\boldmath{$\alpha$}}_{\rm{f}};q_{\rm{i}},
\mbox{\boldmath{$\alpha$}}_{\rm{i}})
\nonumber \\
&&= \frac{w^4}{4\omega_{\rm{i}}\omega_{\rm{f}}}
\left(\frac{2}{15}\right)^2
 \left(\mbox{\boldmath{$\alpha$}}_{\rm{i} \perp}
\times \mbox{\boldmath{$\alpha$}}_{\rm{f} \perp}\right)^2
                  Y^{(1)}(\omega_{\rm{i}};\textrm{q},\omega).
\label{eq.rixs.flip}
\end{eqnarray}
with $\omega\equiv \omega_{\rm{i}}-\omega_{\rm{f}}$ is defined by
\begin{equation}
Y^{(1)}(\omega_{\rm{i}};\textrm{q},\omega)=
\int\langle Z^{(1)\dagger}(\omega_{\rm{i}};\textrm{q},t) 
  Z^{(1)}(\omega_{\rm{i}};\textrm{q},0)\rangle {\rm e}^{\rm{i}\omega t}
{\rm d}t,
\label{eq.y1}
\end{equation}
where 
\begin{eqnarray}
 Z^{(1)}(\omega_{\rm{i}};\textrm{q})&=&
  f_{0}^{(1)}(\omega_{\rm{i}}) 
[S_{a}^{x'}(-\textrm{q})+S_{b}^{x'}(-\textrm{q})]
 \nonumber \\
&+& \textrm{i} \Delta_{\perp,0}^{(1)}(\omega_{\rm{i}})
[S_{a}^{y'}(-\textrm{q})-S_{b}^{y'}(-\textrm{q})].
\end{eqnarray}
Here the time dependent operator of an arbitrary operator $A$ is 
defined as $A(t) = \textrm{e}^{i H_{\rm{mag}} t}A
\textrm{e}^{-i H_{\rm{mag}} t}$.
The Fourier transforms of the spin operators are given by
\begin{eqnarray}
 \textrm{S}_{a}(-\textrm{q})&=& 
(2/N)^{\frac{1}{2}}
   \sum_{i\in \rm{A}}\textrm{S}_{i}\exp(\rm{i}\textrm{q}\cdot \textrm{r}_i),\\
 \textrm{S}_{b}(-\textrm{q})&=& 
(2/N)^{\frac{1}{2}}
   \sum_{i\in \rm{B}}\textrm{S}_{i}\exp(\rm{i}\textrm{q}\cdot \textrm{r}_i),
\end{eqnarray}
where the sum is taken over site $i$ on the A or B sublattices.
The $x'$, $y'$, and $z'$ axes are defined as directing to
$(0,0,1)$, $(1,-1,0)$, and $(1,1,0)$, respectively.
The spin-flip excitations on the neighbouring sites to the core hole are 
neglected, because their amplitudes are quite small.

We expand the spin operators by means of magnon operators in the 
$1/S$-expansion method, which is briefly summarised 
in Appendix \ref{App.D}.
In their expressions, momenta are defined within the first magnetic
Brillouin zone (MBZ). When momentum ${\bf q}$ lies outside the first MBZ,
$\textrm{S}_a(-\textrm{q})$ and $\textrm{S}_b(-\textrm{q})$ are replaced by
$\textrm{S}_a([-\textrm{q}])$ and 
${\rm sgn}(\gamma_\textrm{G})\textrm{S}_b([-\textrm{q}])$, respectively,
where $\textrm{q}$ is put back into the first MBZ by 
a reciprocal lattice vector $\textrm{G}$.
That is, $\textrm{q}=[\textrm{q}]+\textrm{G}$ with $[\textrm{q}]$ lying 
inside the first MBZ.
The ${\rm sgn}(\gamma_{\textrm{k}})$ denotes the sign of $\gamma_{\textrm{k}}$,
where 
$\gamma_{\textrm{k}}=\frac{1}{2}(\cos k_x+\cos k_y)$ with $\textrm{k}$ in units of 
$1/a$ ($a$ is the lattice constant). 
For example, $\gamma_{\textrm{G}}=-1$ for $\textrm{G}=(\pi,\pi)$.
With these notations together with the
magnon operators $\alpha_{[-\textrm{q}]}^{\dagger}$ and 
$\beta_{[-\textrm{q}]}^{\dagger}$, 
$Z^{(1)}(\omega_{\rm{i}};\textrm{q})$ is expressed as
\begin{eqnarray}
 Z^{(1)}(\omega_{\rm{i}};\textrm{q}) &=& 
\sqrt{2S}M(\omega_{\rm{i}};\textrm{q})
\left(\alpha_{[-\textrm{q}]}^{\dagger}
+ {\rm sgn}(\gamma_{\textrm{G}})\beta_{[-\textrm{q}]}^{\dagger}\right)
 + \cdots, \nonumber\\
\end{eqnarray}
where
\begin{eqnarray}
M(\omega_{\rm{i}};\textrm{q}) &\equiv& \frac{\ell_{[\textrm{q}]}}{2}
\Bigl\{
 f_0^{(1)}(\omega_{\rm{i}})
[1-{\rm sgn}(\gamma_{\textrm{G}}) x_{[\textrm{q}]}]
 \nonumber \\
&& +\textrm{i}\Delta_{\perp,0}^{(1)}(\omega_{\rm{i}})
[1+{\rm sgn}(\gamma_{\textrm{G}}) x_{[\textrm{q}]}] \Bigr\}.
\end{eqnarray}
The definitions of $\ell_{\textrm{q}}$ and $x_{\textrm{q}}$ 
are found in Appendix \ref{App.D} [(\ref{eq.xk})] and
use has been made of the relations 
$\ell_{-\textrm{q}}=\ell_{\textrm{q}}$ and $x_{-\textrm{q}}=x_{\textrm{q}}$.
Therefore $Y^{(1)}(\omega_{\rm{i}};\textrm{q},\omega)$ consists of the 
$\delta$-function peak, which is located at 
\begin{equation}
 \omega=JSz\left(1+\frac{A}{2S}\right)\epsilon_{[\textrm{q}]},
\end{equation}
where $A=0.1579$ is the first order correction in the $1/S$-expansion 
(see Appendix \ref{App.D})\cite{Oguchi60}.

Figure \ref{fig.q-dep.one-mag} shows 
$Y^{(1)}(\omega_{\rm{i}};\textrm{q},\omega)$ 
as a function of $\omega$ for $\textrm{q}$ along the symmetry 
directions with $\Gamma/J=2.4$.
A notable aspect is that the intensities diverge 
at $\textrm{q}=(0,0)$ and $(\pi,\pi)$.
The corresponding integrated intensity is given by 
\begin{equation}
I^{(1)}(\omega_{\rm{i}};\textrm{q})\equiv
\int Y^{(1)}(\omega_{\rm{i}};\textrm{q},\omega)
\frac{{\rm d}\omega}{2\pi} 
=2(2S)|M(\omega_{\rm{i}};\textrm{q})|^2.
\end{equation}
Figure \ref{fig.one-mag.int} shows $I^{(1)}(\omega_{\rm{i}};\textrm{q})$ 
for $\textrm{q}$ along symmetry directions with $\Gamma/J=2.4$.
The intensities are enhanced around $\textrm{q}=(0,0)$ as $1/|\textrm{q}|$.
Since the contribution from the isotropic term vanishes around 
$\textrm{q} \sim 0$, 
the enhancement is due to the finite value
of the anisotropic coefficient $\Delta_{\perp,0}^{(1)}(\omega_{\rm{i}})$.
On the other hand, the divergence around $\textrm{q}=(\pi,\pi)$
is brought about by the isotropic term, which is why
the behaviour is irrelevant of presence of the anisotropic term.

\begin{figure}
\includegraphics[width=8.0cm]{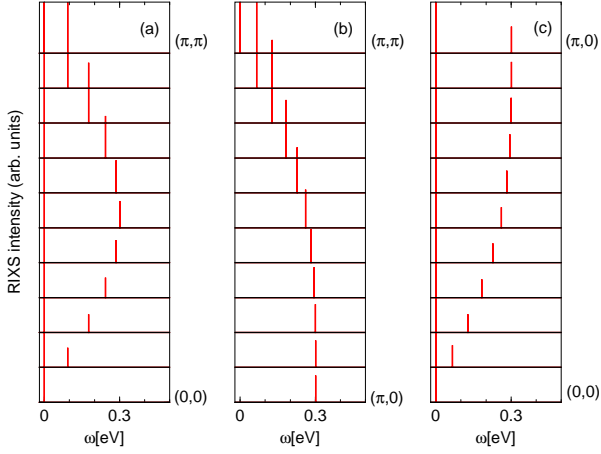}%
\caption{\label{fig.q-dep.one-mag}
$Y^{(1)}(\omega_{\rm{i}};\textrm{q},\omega)$ as a function of $\omega$ 
for $\textrm{q}$ along symmetry directions.
The $\omega_{\rm{i}}$ is set to give rise to the peak in the absorption spectra.
$J=130$ meV and $\Gamma/J=2.4$. The intensities diverge at $\textrm{q}=0$
and $(\pi,\pi)$.
}
\end{figure}

\begin{figure}
\includegraphics[width=8.0cm]{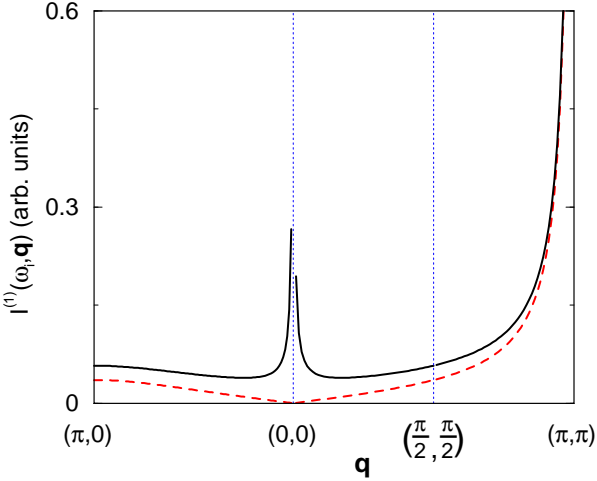}%
\caption{\label{fig.one-mag.int} 
Intensity $I^{(1)}(\omega_{\rm{i}};\textrm{q})$ of the $\delta$-function peak
arising from the one-magnon excitation 
as a function of $\textrm{q}$ along symmetry directions.
The $\omega_{\rm{i}}$ is the photon energy giving rise to the peak 
in the absorption spectra. $\Gamma/J=2.4$.
The (black) solid and (red) broken lines are the results 
with and without including
$\Delta_{\perp,0}^{(1)}(\omega_{\rm{i}})$, respectively.
}
\end{figure}

It has been observed in the RIXS experiments
\cite{Braicovich09,Braicovich10,Guarise10}
that the intensity of magnon peak increases 
significantly with $\textrm{q}\to 0$.
Such increase is consistent with the effects of the anisotropic terms.
So far, the increase of intensity has been interpreted simply as the 
contribution from elastic scattering. To confirm the effects of anisotropic 
terms, it may be necessary to examine carefully the spectra with subtracting
systematically the contribution of elastic scattering 
around $\textrm{q}\sim 0$.

It should be noted here that there exists non-linear terms 
which make the one-magnon excitation split into three-magnon excitations
in the second order correction of the $1/S$-expansion 
\cite{Igarashi92-1,Igarashi2012-1}.
Accordingly $Y^{(1)}(\omega_{\rm{i}};\textrm{q},\omega)$ contains the energy continuum of 
the three-magnon excitations in addition to the $\delta$-function peak
mentioned above.
The contribution from the three-magnon excitation grows gradually
when $\textrm{q}$ is near the boundary of the first MBZ.
See figure 7 in \cite{Igarashi2012-1} for such 
RIXS spectra. 

\subsection{Scattering channel without changing polarization}
In order to calculate the RIXS intensity in this channel,
we collect up the amplitudes from all 
the Cu sites with the use of (\ref{eq.pol-conserve}).
We obtain 
\begin{eqnarray}
&&W(q_{\rm{f}},\mbox{\boldmath{$\alpha$}}_{\rm{f}};q_{\rm{i}},
\mbox{\boldmath{$\alpha$}}_{\rm{i}}) 
\nonumber \\
&=& \frac{w^4}{4\omega_{\rm{i}}\omega_{\rm{f}}}
\left(\frac{2}{15}\right)^2
  \left(\mbox{\boldmath{$\alpha$}}_{\rm{i} \perp} \cdot
\mbox{\boldmath{$\alpha$}}_{\rm{f} \perp}\right)^2                
 Y^{(2)}(\omega_{\rm{i}};\textrm{q},\omega).
\label{eq.rixs.nonflip1}
\end{eqnarray}
The correlation function $Y^{(2)}(\omega_{\rm{i}};\textrm{q},\omega)$ is defined by
\begin{equation}
 Y^{(2)}(\omega_{\rm{i}};\textrm{q},\omega) = \int
 \langle Z^{(2)\dagger}(\omega_{\rm{i}};\textrm{q},t)
Z^{(2)}(\omega_{\rm{i}};\textrm{q},0)\rangle 
  {\rm e}^{\rm{i}\omega t}{\rm d}t,
\label{eq.y2}
\end{equation} 
where
\begin{eqnarray}
 Z^{(2)}(\omega_{\rm{i}};\textrm{q})&=&
   f_2^{(2)}(\omega_{\rm{i}})[(\textrm{S}\cdot \textrm{X})_{a}(-\textrm{q})
                           +(\textrm{S}\cdot \textrm{X})_{b}(-\textrm{q})]
\nonumber\\
  &+&\Lambda^{(2)}(\omega_{\rm{i}})[(S^{z'}X^{z'})_{a}(-\textrm{q})
                            +(S^{z'}X^{z'})_{b}(-\textrm{q})]\nonumber\\
&+& \cdots.
\end{eqnarray}
The Fourier transform $(S^{\mu '}X^{\mu '})_a(-\textrm{q})$ and 
$(S^{\mu '}X^{\mu '})_b(-\textrm{q})$ for $\mu '=x'$, $y'$, and $z'$
are introduced as follows.
\begin{eqnarray}
&& \hspace*{-0.5cm} (S^{\mu '}X^{\mu '})_a(-\textrm{q}) =
(2/N)^{\frac{1}{2}}\sum_{i\in \rm{A}}
 S_i^{\mu '}\frac{1}{4}\sum_{\delta}S_{i+\delta}^{\mu '}
   \textrm{e}^{\rm{i} \textrm{q}\cdot \textrm{r}_{i}} \nonumber\\
 &=&(2/N)^{\frac{1}{2}}\sum_{\textrm{k}}S_a^{\mu '}(\textrm{k})
S_b^{\mu '}(-[\textrm{k}+\textrm{q}])
   \gamma_{[\textrm{k}+\textrm{q}]} , \\
&& \hspace*{-0.5cm}(S^{\mu '}X^{\mu '})_b(-\textrm{q}) 
= (2/N)^{\frac{1}{2}}\sum_{i\in \rm{B}}
 S_i^{\mu '}\frac{1}{4}\sum_{\delta}S_{i+\delta}^{\mu '}
   \textrm{e}^{\rm{i} \textrm{q}\cdot \textrm{r}_{i}} \nonumber\\
 &=&(2/N)^{\frac{1}{2}}\sum_{\textrm{k}}S_b^{\mu '}(\textrm{k})
S_a^{\mu '}(-[\textrm{k}+\textrm{q}])
   \gamma_{[\textrm{k}+\textrm{q}]}{\rm sgn}(\gamma_{\textrm{G}}). 
\nonumber \\
\end{eqnarray}
Here the sum over $\delta$ is carried out on the nearest neighbour 
sites around site $i$. 

Expanding $Z^{(2)}(\omega_{\rm{i}};\textrm{q})$ in terms of magnon operators
within the $1/S$-expansion(see Appendix \ref{App.D}), we obtain
\begin{equation}
  Z^{(2)}(\omega_{\rm{i}};\textrm{q})=
(2S)\sum_{\textrm{k}}N(\omega_{\rm{i}};\textrm{q},\textrm{k})
           \alpha_{-[\textrm{q}+\textrm{k}]}^{\dagger}\beta_{\textrm{k}}^{\dagger}
           + \cdots,
\label{eq.z2}
\end{equation}
with $\textrm{k}$ running within the first MBZ, and    
\begin{eqnarray}
 && N(\omega_{\rm{i}};\textrm{q},\textrm{k}) \nonumber\\
& =& f_2^{(2)}(\omega_{\rm{i}})\frac{\ell_{[\textrm{q}+\textrm{k}]}
\ell_{\textrm{k}}}{2} 
 \left\{ - (1+\gamma_{\textrm{q}})\left[x_{\textrm{k}}
  +{\rm sgn}(\gamma_{\textrm{G}}) x_{[\textrm{q}+\textrm{k}]}\right]
\right. \nonumber \\
&&\left. +\left[\gamma_{\textrm{k}}+{\rm sgn}
(\gamma_{\textrm{G}})\gamma_{[\textrm{q}+\textrm{k}]}\right] 
[ 1 + {\rm sgn}(\gamma_{\textrm{G}}) x_{[\textrm{q}+\textrm{k}]}x_{\textrm{k}}]
\right\}
\nonumber\\
 &-& 
 \Lambda^{(2)}(\omega_{\rm{i}})
\frac{\ell_{[\textrm{q}+\textrm{k}]}\ell_{\textrm{k}}}{2}
 (1+\gamma_{\textrm{q}})\left[x_{\textrm{k}}
  +{\rm sgn}(\gamma_{\textrm{G}}) x_{[\textrm{q}+\textrm{k}]}\right].
\end{eqnarray}
This expression is valid even when
$\textrm{q}$ is outside of the first MBZ.
Note that when $\Lambda^{(2)}(\omega_{\rm{i}})=0$, 
$N(\omega_{\rm{i}};\textrm{q},\textrm{k})$ vanishes 
at $\textrm{q}=(0,0)$ and $(\pi,\pi)$\cite{Nagao07}.
Note also that the isotropic terms of the two-magnon part 
are the same as those obtained for the $K$-edge RIXS, 
where no anisotropic term exists\cite{Nagao07,Hill08,Forte08}.

\begin{figure}
\includegraphics[width=8.0cm]{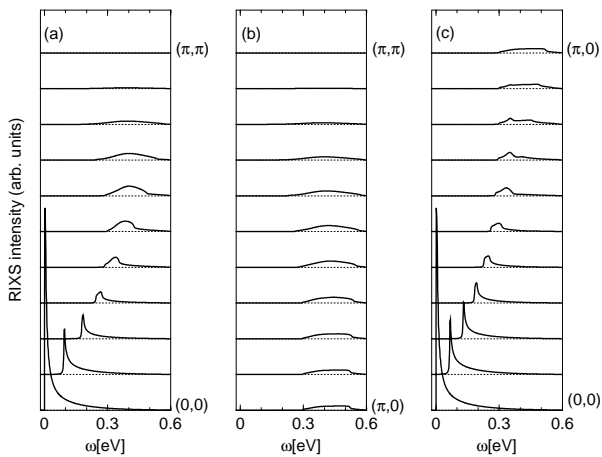}%
\caption{\label{fig.q-dep.two-mag}
$Y^{(2)}(\omega_{\rm{i}};\textrm{q},\omega)$ as a function of $\omega$ 
for $\textrm{q}$ along symmetry directions.
The $\omega_{\rm{i}}$ is set to give rise to the peak in the absorption spectra.
$J=130$ meV and $\Gamma/J=2.4$.
}
\end{figure}

From (\ref{eq.z2}), we see that 
$Y^{(2)}(\omega_{\rm{i}};\textrm{q},\omega)$
consists of the energy continuum of the two-magnon excitations.
Since two magnons are created at neighbouring sites through x-ray scattering, inclusion of 
the magnon-magnon interaction is crucial to obtain the spectral shape.
As already discussed in \cite{Nagao07}, 
the magnon-magnon interaction in the $1/S$-expansion 
could be divided into a separable form so that the $t$-matrix of the 
scattering is neatly evaluated.  We resort to the similar evaluation. 
Figure \ref{fig.q-dep.two-mag} shows 
$Y^{(2)}(\omega_{\rm{i}};\textrm{q},\omega)$ 
as a function of $\omega$ for $\textrm{q}$ along the symmetry directions.
We find rapid enhancement of the intensity is brought about by 
the presence of the anisotropic terms as 
$|\textrm{q}|$ goes to $(0,0)$. 
Without them, in contrast, the intensity diminishes in this limit
as shown in figure 8 of \cite{Igarashi2012-1}.
We see the peak energy decreases with $|\textrm{q}|$ approaching zero. 
At $\textrm{q}=0$, the peak energy becomes very close to zero, 
$\sim 0.025$eV. 
It may be a difficult task to distinguish the spectral peak from the 
elastic peak.
However a careful study on the $\textrm{q}$-dependence of the spectra may clarify
such effect of the anisotropic terms.

\begin{figure}
\includegraphics[width=8.0cm]{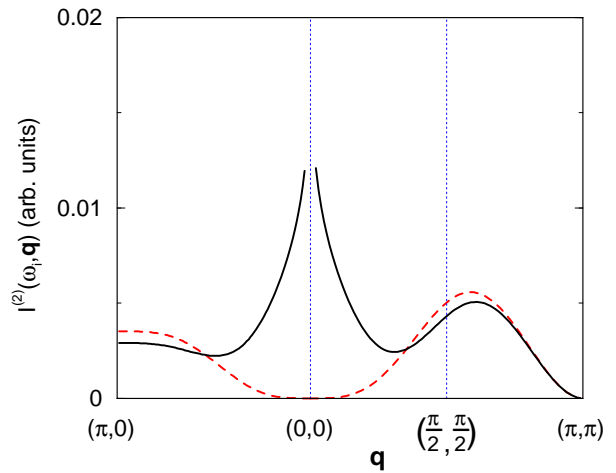}%
\caption{\label{fig.two-mag}
Frequency-integrated intensity $I^{(2)}(\omega_{\rm{i}};\textrm{q})$ of the continuous 
spectra arising from two-magnon excitations
as a function of $\textrm{q}$ along symmetry directions.
The $\omega_{\rm{i}}$ is the photon energy giving rise to the peak 
in the absorption spectra. $\Gamma/J=2.4$.
The (black) solid and (red) broken lines represent
the results obtained with and without the anisotropic term,
respectively.
}
\end{figure}

The frequency integrated intensity $I^{(2)}(\omega_{\rm{i}};\textrm{q})$
may be given by
\begin{eqnarray}
 I^{(2)}(\omega_{\rm{i}};\textrm{q}) &\equiv& 
\int Y^{(2)}(\omega_{\rm{i}};\textrm{q},\omega)
 \frac{{\rm d}\omega}{2\pi} 
\nonumber\\
 &=& (2S)^2\sum_{\textrm{k}}
\left| N(\omega_{\rm{i}};\textrm{q},\textrm{k})\right|^2.
\end{eqnarray}
Figure \ref{fig.two-mag} shows $I^{(2)}(\omega_{\rm{i}};\textrm{q})$ 
for $\textrm{q}$ along symmetry directions. 
Notice that $I^{(2)}(\omega_{\rm{i}};\textrm{q})$
diverges logarithmically when $|\textrm{q}|$ approaches zero.
It demonstrates a possibility 
that this logarithmic enhancement can be
recognized at a region where $\textrm{q}$ is away from $(0,0)$.
 
\subsection{Polarization dependence}

\begin{figure}
\includegraphics[width=8.0cm]{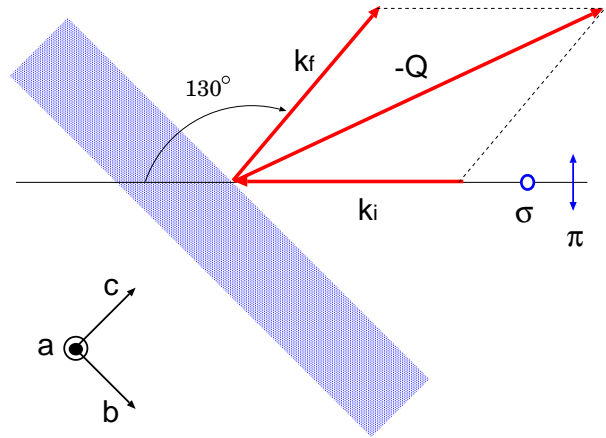}
\caption{\label{fig.scat.geo}
Schematic view of the scattering geometry.
The angle between the incident and scattered x-rays is fixed at 130 degrees.
The scattering plane contains the $c$ axis.
}
\end{figure}

We consider a scattering geometry used in the experiments of La$_2$CuO$_4$
\cite{Braicovich10} and Sr$_2$CuO$_2$Cl$_2$\cite{Guarise10}. 
It is schematically shown 
in figure \ref{fig.scat.geo} for $\textrm{q}$ along $(0,0)-(0,\pi)$ direction,
where the angle between the incident and the scattered x-ray 
is fixed at 130 
degrees. The scattering plane includes the $b(x)$ and $c(z)$ axes.

Then $\mbox{\boldmath{$\alpha$}}_{\rm{i}\perp}=(1,0,0)$ 
for the $\sigma$ polarization and
$\mbox{\boldmath{$\alpha$}}_{\rm{i}\perp}=(0,\alpha_{\rm{i}}^{\pi},0)$ 
for the $\pi$ polarization in the incident x-ray, while 
$\mbox{\boldmath{$\alpha$}}_{\rm{f}\perp}=(1,0,0)$ for the $\sigma'$ polarization 
and $\mbox{\boldmath{$\alpha$}}_{\rm{f}\perp}
=(0,\alpha_{\rm{f}}^{\pi},0)$ 
for the $\pi'$ polarization in the scattered x-ray. 
Thereby the RIXS spectra may be expressed as
\begin{eqnarray}
&&W(q_{\rm{f}},\mbox{\boldmath{$\alpha$}}_{\rm{f}};q_{\rm{i}},
\mbox{\boldmath{$\alpha$}}_{\rm{i}})  \nonumber \\
&&= 
\frac{w^4}{4\omega_{\rm{i}}\omega_{\rm{f}}}
\left(\frac{2}{15}\right)^2 
\times 
\left\{ \begin{array}{ll}
 (\alpha_{\rm{f}}^{\pi})^2 
Y^{(1)}(\omega_{\rm{i}};\textrm{q},\omega), & \sigma\to\pi \\
 Y^{(2)}(\omega_{\rm{i}};\textrm{q},\omega), & \sigma\to\sigma'\\
 (\alpha_{\rm{i}}^{\pi})^2 Y^{(1)}(\omega_{\rm{i}};\textrm{q},\omega), & \pi\to\sigma'\\
 (\alpha_{\rm{f}}^{\pi}\alpha_{\rm{i}}^{\pi})^2 
Y^{(2)}(\omega_{\rm{i}};\textrm{q},\omega),
  &  \pi\to\pi'
\end{array} \right. .\nonumber\\
\end{eqnarray}
The one-magnon term $Y^{(1)}$ and the two-magnon term $Y^{(2)}$ 
are separated by the polarization.
Accordingly the polarization analysis is useful to clarify the contribution
of $Y^{(2)}$. For other directions of $\textrm{q}$, we could obtain the similar 
formulas separated by polarizations.

\section{\label{sect.6} Concluding Remarks}

We have studied the magnetic excitations on the $L$-edge RIXS 
from undoped cuprates beyond the FCA.
Emphasis is on how the symmetry breaking of the ground state affects
the magnetic RIXS spectra.
It is found that the spin excitations are brought about 
at neighbouring sites in addition to the core-hole site.
We have shown that the anisotropic terms emerged in the scattering
amplitudes as a direct consequence of the broken symmetry.
The fact contrasts sharply with the case in neutron
scattering, where the amplitude is described through 
the interaction Hamiltonian between the spins of 
neutron and electron. The presence of such anisotropic terms has been 
supported by the calculation on a one-dimensional finite-size ring of spins 
under the staggered external field and on a two-dimensional cluster with the 
molecular field acting on the boundary. 
Collecting up such amplitudes on all the Cu sites,
we have expressed the RIXS spectra in the form of spin correlation functions,
which have been calculated within the $1/S$-expansion. The anisotropic terms
have made the RIXS intensity considerably enhanced as $|\textrm{q}|$ goes to zero.
Such enhancement could be confirmed experimentally by observing carefully
the spectra around $\textrm{q}=0$.
With a little further improvement on energy resolution, it
would make the
distinction possible as achieved in Sr$_2$IrO$_4$,
in which band-splitting, predicted by a theory\cite{Igarashi2013-1}, 
conspicuous around $\textrm{q}=(0,0)$ had been discerned 
by recent experiment\cite{JKim2014}.
We believe our present emphasize on the anisotropic terms 
originated from the
antiferromagnetic long range order might be insightful when
one analyses the systems with short range order 
such as the doped high-$T_{\rm{c}}$ cuprates
\cite{Chen2013,Jia2014,Benjamin.pp}.

\section{acknowledgments}
This work was partially supported by a Grant-in-Aid for Scientific Research 
from the Ministry of Education, Culture, Sports, Science and Technology
of the Japanese Government.

\appendix
\section{Absorption coefficient\label{App.A}}
Since the $2p$ core hole is quite localized in real space,
the absorption coefficient is well approximated by the sum
of the intensities on each lattice site.
Therefore, after averaging the polarization, 
the absorption coefficient $A(j ,\omega_{\rm{i}})$ 
at the Cu $L_{2}$- and $L_3$-edges may be expressed as
\begin{equation}
 A(j ,\omega_{\rm{i}}) \propto  \frac{\Gamma}{\pi}
  \sum_{\sigma,\eta}
\frac{|\langle \phi_{\eta}|\psi_0^{\sigma}\rangle|^2}
     {[\omega_{\rm{i}}+\epsilon_{\rm{g}}-\epsilon_{\rm{core}}
  -\epsilon'_{\eta}]^2+\Gamma^2},
\label{eq.abs}
\end{equation}
where $\epsilon_{\rm core}$ depends on $j$.
By substituting the eigenvalues and the eigenstates evaluated on finite-size
clusters into (\ref{eq.abs}), we obtain $A(j,\omega_{\rm{i}})$.
For $\Gamma$ comparable or larger than the excitation energy 
$\epsilon'_{\eta}$, $A(j,\omega_{\rm{i}})$ is close to the Lorentzian curve.
The calculated $A(j,\omega_{\rm{i}})$ for 
$\Gamma/J=2.4$ in a two-dimensional 
Heisenberg model has been shown in figure 3 in \cite{Igarashi2012-1}.

\section{Projection onto non-orthogonal bases\label{App.B}}
We try to project a state $|F\rangle $ onto the
non-orthogonal states $|\Psi_i\rangle$'s as
\begin{equation}
 |F\rangle = \sum_{i}f_i |\Psi_i\rangle.
\label{eq.proj}
\end{equation}
Operating $\langle\Psi_i|$ from the left side of
(\ref{eq.proj}), we obtain
$Q_i=\sum_{j}\hat{\rho}_{i,j}f_j$ where 
$Q_i\equiv \langle\Psi_i|F\rangle$
and $\hat{\rho}_{i,j}\equiv \langle\Psi_i|\Psi_j\rangle$.
Therefore the expansion coefficients $f_i$'s are given by
$f_i=\sum_{j}(\hat{\rho}^{-1})_{i,j}Q_{j}$.
As long as the number of the projected states remains finite,
this procedure uniquely determines the expansion coefficients.

\section{One-dimensional ring of spins under the staggered external field
\label{App.C}}
\begin{figure}
\includegraphics[width=8.0cm]{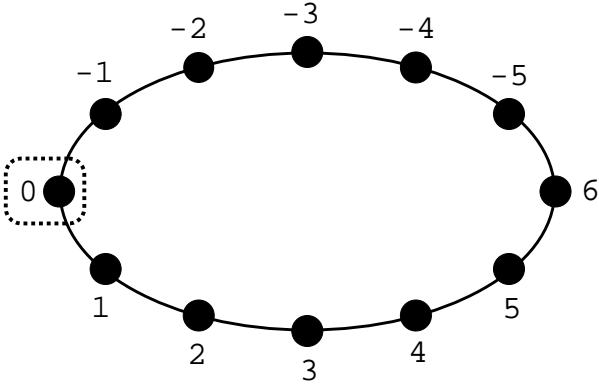}
\caption{\label{fig.ring}
A ring of 12 spins ($S=\frac{1}{2}$) used to evaluate 
the coefficients for the spin
excitations in RIXS.
The spin at site 0 is annihilated in the intermediate state.
}
\end{figure}

We examine how the anisotropic terms develop in concert with the development
of the staggered moment. This is achieved easily in one-dimensional
system, since there is no long range order in the ground state. 
To control the staggered moment, we apply the staggered external field. 
Thereby the Hamiltonian of the system is given by
\begin{equation}
 H_{\rm mag} = J\sum_{\langle i,j\rangle}\textrm{S}_i\cdot \textrm{S}_j
  + H_{\rm ex}\sum_{i}(-1)^{i}S_{i}^{z'},
\end{equation}
where the field strength is denoted as $H_{\textrm{ex}}$.
We consider a system made of 12 spins of $S=\frac{1}{2}$ with periodic 
boundary condition, as shown in figure \ref{fig.ring}.

\begin{figure}
\includegraphics[width=8.0cm]{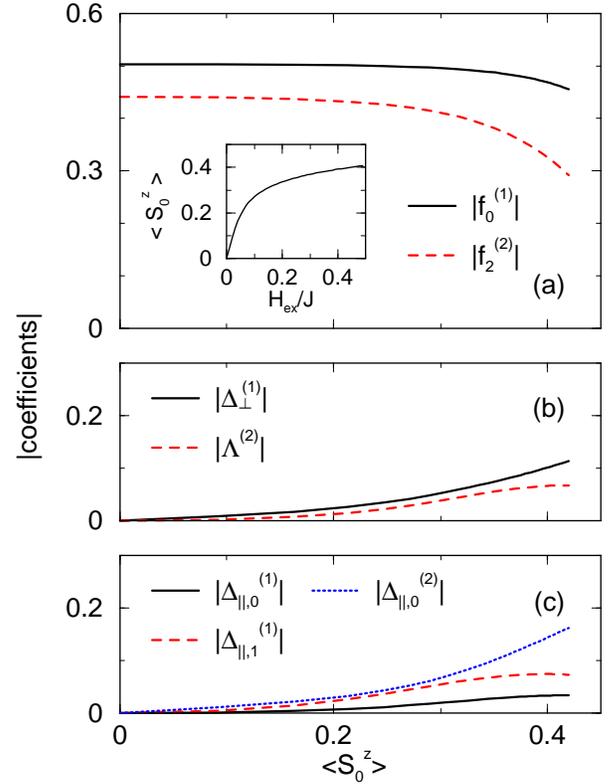}
\caption{\label{fig.1d.coef}
Various coefficients measured in units of $J^{-1}$ 
as a function of the staggered moment
evaluated in a ring of 12 spins. $\Gamma/J=2.0$, and $\omega_{\rm{i}}$ 
is set to give rise to the maximum absorption coefficient. 
(a) The (black) solid and (red) broken curves correspond
to $|f_0^{(1)}|$ and $|f_2^{(2)}|$, respectively.
(b) The (black) solid and (red) broken curves represent
$|\Delta_{\perp}^{(1)}|$ and $|\Lambda^{(2)}|$, respectively.
(c) The (black) solid, (red) broken, and (blue) dotted curves show
$|\Delta_{\parallel,0}^{(1)}|$, 
$|\Delta_{\parallel,1}^{(1)}|$, and 
$|\Delta_{\parallel,0}^{(2)}|$, respectively.
The inset in Panel (a) shows the staggered moment as a function of 
$H_{\rm ex}/J$.
}
\end{figure}

Since the total-spin component along the $z'$ axis is zero 
in the ground state, the Hamiltonian matrix is represented 
by a matrix with $924\times 924$ dimensions. 
Diagonalizing the Hamiltonian matrix, we obtain
the ground-state wavefunction. The inset in figure \ref{fig.1d.coef} (a)
shows the staggered magnetization 
$\langle S_{0}^{z'}\rangle$ as a function of $H_{\rm ex}$, 
which increases with increasing $H_{\rm ex}$.
The intermediate state, on the other hand, consists of 11 spins,
since the spin degree of freedom is lost at the core-hole site. 
Thereby the Hamiltonian matrix in the intermediate state is 
represented by a 
matrix with $462\times 462$ dimensions in the subspace of 
$\sum_{i}S_{i}^{z '}=\pm \frac{1}{2}$.
Using the eigenvalues and eigenfunctions for the Hamiltonian of the 
intermediate state together with the ground state, we evaluate
(\ref{eq.proj1}), (\ref{eq.proj2}), and others for the coefficients of 
the spin excitations.

\begin{table}
\caption{\label{table.2}
Coefficients for isotropic terms in units of $1/J$ in the one-dimensional
ring of 12 spins ($S=\frac{1}{2}$). The incident photon energy
$\omega_{\rm{i}}$ is set to give the maximum absorption coefficient.
}
\footnotesize
\begin{tabular}{cccc}
\hline
\hline
 $\Gamma/J$ & $f_{0}^{(1)}(\omega_{\rm{i}})$ & 
               $f_{1}^{(1)}(\omega_{\rm{i}})$ & 
               $f_2^{(2)}(\omega_{\rm{i}})$ \\
\hline
       $2.0$ & $(-0.013,-0.500)$ & $(0.043,-0.025)$ & $(0.366,-0.219)$ \\
       $1.5$ & $(0.025,-0.672)$ & $(0.088,-0.051)$ & $(0.633,-0.365)$  \\
       $1.0$ & $(0.050,-1.009)$ & $(0.154,-0.125)$ & $(1.101,-0.894)$  \\
\hline
\hline
\end{tabular}
\end{table}
\normalsize

By setting $H_{\textrm{ex}}=0$, we first evaluate the 
isotropic terms in the absence of the anisotropic terms.
Table \ref{table.2} shows the coefficients for isotropic terms 
for several values of $\Gamma/J$ with 
$\omega_{\rm{i}}$ being fixed
at the value to give the maximum absorption coefficient.
The values $\Gamma/J=2.0$ and $1.5$ may correspond to 
CaCu$_2$O$_3$\cite{Lake2010} and 
Sr$_2$CuO$_3$\cite{Walters2009}, respectively.
The coefficient $f_{1}^{(1)}(\omega_{\rm{i}})$ for the 
spin-flip excitation 
on neighbouring sites is much smaller than $f_{0}^{(1)}(\omega_{\rm{i}})$.
The coefficient $f_2^{(2)}(\omega_{\rm{i}})$ for the 
$\textrm{S}_0\cdot \textrm{X}$ term
is comparable to the coefficient $f_0^{(1)}(\omega_{\rm{i}})$ 
for the spin-flip
term. It grows with decreasing $\Gamma/J$, as was discussed in 
\cite{Igarashi2012-2}.

Next, we turn our attention to the anisotropic terms. 
Figures \ref{fig.1d.coef}(b) and (c)
show the absolute values of the coefficients 
as a function of staggered moment for $\Gamma/J=2.0$.
They demonstrate that the anisotropic terms develop with increasing
staggered moment as expected. 
Note that the magnitudes of the isotropic terms vary 
gradually and slightly diminish rather than increase
with increasing staggered moment
as shown in figure \ref{fig.1d.coef} (a).

\section{$1/S$-expansion\label{App.D}}
Here, we briefly summarise an introduction of the $1/S$ expansion.
The emphasis is on the definitions of the quantities used in the main
text. The details are relegated to the references such as 
\cite{Igarashi2012-1}.
Assuming two sublattices in the antiferromagnetic ground state, 
we express spin operators by boson operators as\cite{Holstein40}
\begin{eqnarray}
 S_i^z &=& S - a_i^\dagger a_i ,  
 \label{eq.boson1}\\
 S_i^+ &=& (S_i^-)^\dagger = \sqrt{2S}f_i(S)a_i , \\
 S_j^z &=& -S + b_j^\dagger b_j ,\\
 S_j^+ &=& (S_j^-)^\dagger = \sqrt{2S}b_j^\dagger f_j(S) ,
 \label{eq.boson2}
\end{eqnarray}
where $a_i$ and $b_j$ are boson annihilation operators, and
\begin{equation}
    f_\ell (S) = \sqrt{1 - \frac{n_\ell}{2S}}
               = 1 - \frac{1}{2}\frac{n_\ell}{2S} -
                  \frac{1}{8}\left(\frac{n_\ell}{2S}\right)^2 + \cdots ,
\end{equation}
with $n_\ell$ represents $a_i^\dagger a_i$ and $b_j^\dagger b_j$ for 
$\ell=i$ and $j$, respectively.
Indices $i$ and $j$ refer to sites on the A and B 
sublattices, 
respectively. Using (\ref{eq.boson1})-(\ref{eq.boson2}),
$H_{\rm mag}$ is expanded in powers of $1/S$,
\begin{equation}
 H_{\rm mag} = -JS^2Nz/2 + H_{\rm mag}^{(0)} + H_{\rm mag}^{(1)} 
+ \cdots, 
\end{equation}
where $N$ and $z$ are the number of lattice sites and that of nearest neighbour
sites, respectively. $H_{\rm{mag}}^{(n)}$ stands
for the $n$-th order term in the $1/S$-expansion.
The Fourier transforms of the boson operators are introduced 
within the first MBZ,
\begin{equation}
 a_i = (2/N)^{\frac{1}{2}}\sum_{\textrm{k}} a_{\textrm{k}}
  \textrm{e}^{\rm{i} \textrm{k}\cdot \textrm{r}_i},
 b_j = (2/N)^{\frac{1}{2}}
       \sum_{\textrm{k}} b_{\textrm{k}} 
\textrm{e}^{\rm{i} \textrm{k}\cdot \textrm{r}_j}.
\end{equation}
Then, with the help of a Bogoliubov transformation,
\begin{equation}
 a_{\textrm{k}}^\dagger 
= \ell_{\textrm{k}}\alpha_{\textrm{k}}^\dagger
          +m_{\textrm{k}}\beta_{-\textrm{k}}, \quad
 b_{-\textrm{k}} = 
m_{\textrm{k}}\alpha_{\textrm{k}}^\dagger
          + \ell_{\textrm{k}}\beta_{-\textrm{k}}, 
\label{eq.magnon}
\end{equation}
we could diagonalize $H_{\rm mag}^{(0)}$ as
\begin{equation}
 H_{\rm mag}^{(0)} = JSz\sum_{\textrm{k}}(\epsilon_{\textrm{k}}-1) 
     + JSz\sum_{\textrm{k}} \epsilon_{\textrm{k}}
   (\alpha_{\textrm{k}}^\dagger \alpha_{\textrm{k}}
   + \beta_{\textrm{k}}^\dagger\beta_{\textrm{k}}).
\end{equation}
Here,
\begin{eqnarray}
 \ell_{\textrm{k}} &=& \sqrt{\frac{1+\epsilon_{\textrm{k}}}
 {2\epsilon_{\textrm{k}}}}, \ \
  m_{\textrm{k}} =- 
\sqrt{\frac{1-\epsilon_{\textrm{k}}}{2\epsilon_{\textrm{k}}}}
= -x_{\textrm{k}} \ell_{\textrm{k}},
\label{eq.xk}\\
 \epsilon_{\textrm{k}} &=& \sqrt{1-\gamma_{\textrm{k}}^2}, \ \
 \gamma_{\textrm{k}} = \frac{1}{z}\sum_{\mbox{\boldmath{$\delta$}}}
     \exp(\rm{i} \textrm{k} \cdot \mbox{\boldmath{$\delta$}}),
\label{eq.gam}
\end{eqnarray}
where $\mbox{\boldmath{$\delta$}}$ connects the origin with the nearest 
neighbour sites.
The expression for $H_{\rm mag}^{(1)}$
becomes\cite{Harris71}
\begin{eqnarray}
&& \hspace*{-0.5cm} H_{\rm mag}^{(1)} 
= \frac{JSz}{2S} A\sum_{\textrm{k}}\epsilon_{\textrm{k}}
 (\alpha_{\textrm{k}}^\dagger \alpha_{\textrm{k}}
+ \beta_{\textrm{k}}^\dagger\beta_{\textrm{k}})
 \nonumber \\
     &+&\frac{-JSz}{2SN}
    \sum_{\textrm{k}_1, \textrm{k}_2, \textrm{k}_3,\textrm{k}_4}
    \delta_{\textrm{G}}(\textrm{k}_1 + \textrm{k}_2 - \textrm{k}_3 -\textrm{k}_4)
\nonumber \\
     &\times& 
\ell_{\textrm{k}_1}\ell_{\textrm{k}_2}\ell_{\textrm{k}_3} \ell_{\textrm{k}_4}
\left(
4\alpha_{\textrm{k}_1}^\dagger\beta_{-\textrm{k}_4}^\dagger
\beta_{-\textrm{k}_2}\alpha_{\textrm{k}_3} B_{\textrm{k}_1 \textrm{k}_2 \textrm{k}_3
\textrm{k}_4}^{(3)} +
\cdots
\right),\nonumber\\
\label{eq.intham}
\end{eqnarray}
with $A=\frac{2}{N}\sum_{\textrm{k}}(1-\epsilon_{\textrm{k}})$\cite{Oguchi60}.
For the square lattice, $A=0.1579$. The Kronecker delta 
$\delta_{\textrm{G}}(\textrm{k}_1 + \textrm{k}_2 - \textrm{k}_3 -\textrm{k}_4)$
indicates the conservation of momenta within a reciprocal lattice vector 
$\textrm{G}$. 
In the second term of (\ref{eq.intham}), only  
the relevant term representing scattering of two magnons 
is shown explicitly.
The vertex function $B^{(3)}$ in a symmetric parametrization
as well as omitted terms are found 
in \cite{Igarashi92-1,Harris71,Syromyatnikov2010}.

\bibliographystyle{apsrev}
\bibliography{paper}

\begin{thebibliography}{40}
\bibitem[{Com()}]{Com1}
\bibinfo{note}{Notice that, in our previous paper \cite{Igarashi2012-1}, the
  symbol $f_n^{(1)}(\omega_{\rm i})$ instead of $g_n^{(1)}(\omega_{\rm i})$ was
  used since they are equivalent when the anisotropic terms are neglected or in
  the absence of long-range order.}
\end{thebibliography}

\end{document}